# Automating Requirements Traceability: Two Decades of Learning from KDD

Alex Dekhtyar
Department of Computer Science and Software Engineering
Cal Poly, San Luis Obispo
dekhtyar@calpoly.edu

Jane Huffman Hayes
Department of Computer Science
University of Kentucky
hayes@cs.uky.edu

*Abstract*—This paper summarizes our experience with using Knowledge Discovery in Data (KDD) methodology for automated requirements tracing, and discusses our insights.

## I. INTRODUCTION

Next year will mark the 20th anniversary of the pioneering work of Antoniol et al. on automating the requirements tracing process [1]. This work was followed closely by Marcus and Maletic [18] and Hayes, Dekhtyar, and Sundaram [15]. As a result, the idea of bringing techniques from the Information Retrieval, and later the Machine Learning, communities[1] to automate requirements tracing has found wide spread acceptance in the Requirements Engineering community, and has yielded a variety of important developments in the past 18 years [17, 5].

In this paper we summarize the experiences of using Knowledge Discovery in Data (KDD) techniques for automating requirements tracing, and share our thoughts on the benefits we reaped and the challenges we faced.

## II. USING KDD TECHNIQUES FOR AUTOMATED REQUIREMENTS TRACING

The requirements tracing problem [13] can be formulated as follows: given a collection of software requirements and another artifact of the software engineering process (design document, code, use cases, test cases, bug reports, etc...): find, for each requirement, individual elements in the other artifact that relate to it.

Tracing, "the activity of either establishing or using traces" [7], is the solution to providing information about the relationships between different artifacts of the software development lifecycle. Such information contributes to (Independent) Validation & Verification of mission- and safety-critical systems, supports change impact analysis, criticality analysis, regression testing, and other important activities.

In this paper, we understand the term *trace* as "the act of following a trace link from a source artifact to a target artifact (primary trace link direction) or vice-versa (reverse trace link direction" [7]. A *trace link* is defined as "a specified association between a pair of artifacts, one comprising the source artifact and one comprising the target artifact." [7]. *Trace matrices*, otherwise known as *trace relations* are collections of vetted (validated) trace links between a pair of artifacts. A *candidate link* is trace link that has not been vetted (i.e., pronounced valid), while a *candidate trace matrix* is a collection of candidate links. In empirical studies related to tracing, we often refer to *ground truth* or *gold standard* trace relations, or *answer sets*: the trace relations determined ahead of time to be correct that have to be captured by whatever tracing methodology is under investigation.

To automate tracing, the pioneering works [1, 18, 15] adopted the Information Retrieval approach of (a) representing the requirements and the elements of other artifacts as vectorized collections of features, and (b) determining related pairs by computing a formal similarity score over the vectorized representations. A wide range of approaches including Naïve Bayes (a.k.a. Probabilistic Information Retrieval) [21], tf-idf Vector Space retrieval [20] and Latent Semantic Indexing (LSI) [10] have been studied at the outset. Later, additional techniques, incorporating a wider range of Machine Learning algorithms, such as, Latent Dirichlet Allocation (LDA) [3] feedback analysis [22], and more were used to various degrees of success over the past 15 years [14, 19, 6]. For a more thorough overview of the work on automated traceability we refer the reader to the surveys of Winkler and von Pilgrim [23] and Borg et al. [4].

In what follows, we summarize both our own experience of adopting and adapting KDD techniques for automated tracing, as well as the experiences of our colleagues shared with us at a variety of fora including, but not limited to, the biannual TEFSE (Traceability in Emerging Forms of Software Engineering) workshops (renamed in 2015 to International Symposium on Software and Systems Traceability), and the 2007 and 2017 Symposia on Grand Challenges in Traceability (GCT'2007 and GCT'2017).

We make the following core observations summarizing our collective experiences.

*a) Tracing is a "small data" problem:* The term "big data" has become popular in recent years, and the vast majority of KDD techniques have been addressing the need to derive insight from massive collections of data (such as millions of email messages, billions of business transactions, or petabytes of scientific observation data). Many Machine Learning and Information Retrieval techniques specifically rely on the abundance of data to construct classification models, or representations of data points.

---
[1] In this paper we use the term "KDD"="Knowledge Discovery in Data" to refer to the overarching field of intelligent information analysis which includes both Information Retrieval and Machine Learning as its constituent parts.

Compared to such problems, requirements tracing is a *feature-poor small data* problem. It is feature-poor because individual requirements, design elements, code components, bug reports, test cases, etc., are relatively short documents - typically the equivalent of about a paragraph (often – a short one) of text. It is a small data problem because in a typical requirements tracing scenario the total number of data points (individual requirements, design elements, test cases, etc.) is relatively small - on the order of tens, sometimes hundreds of individuals.

There are important implications of this fact. KDD methods proven to be robust in big data settings may not necessarily yield more accurate trace relationships simply because these methods do not receive enough information to learn.

*b) Measurements of Success:* The automated requirements tracing community fully adopted both the primary measures of accuracy used in KDD: precision (percent of retrieved answers that are correct), recall (percent of correct answers that were retrieved), f-measure (a possibly weighted harmonic mean of precision and recall), as well as the secondary measures: Mean Average Precision (MAP)[2], area under the ROC[3] curve [12], etc. to evaluate the quality of trace relationships. This adoption of KDD measures *transcended their original intended use* for evaluation of the quality of automated tracing techniques and is now used to evaluate trace relationships regardless of how they were constructed: automatically, manually, or through a hybrid approach.

*c) Relative importance of accuracy measures:* Tracing is an example of what Berry calls a "hairy task" [2]: a task where detecting a false positive is much simpler and faster than discovering an error of omission. As such, when evaluating the results of automated methods generating candidate traces, it is important to understand that *recall*, i.e., the percentage of true links discovered by the tracing method, is *significantly more important* than *precision*, the percentage of discovered links that are true. This is in direct contrast to how Information Retrieval methods are evaluated on typical tasks, such as web search, where precision — i.e., lack of false positive links on the returned page of links — is more important than recall[4].

*d) Lack of datasets is stifling:* KDD research and development gave rise to large dataset collections: from the UCI Machine Learning repository[5], to TREC dataset collections[6], to Kaggle[7]. Yet there is a distinct dearth of good datasets for requirements tracing. This has to do with two major obstacles: (a) difficulty of converging on the ground truth, and (b) the overall lack of publicly available high-quality trace relations for large projects.

*e) No one believes our ground truth:* Another major issue we have encountered in our work on tracing projects is the distinct suspicion with which the larger Requirements Engineering community views ground truth produced as part of tracing research. This is in contrast with the high degree of trust the KDD community generally puts in the ground truth for its datasets. The amount of correspondence with reviewers, and the amount of text that had to be included in the papers (only by the authors, not to mention similar efforts by their colleagues) explaining the origins of the ground truth trace relations far exceeds what one does in KDD or Data Science communities.

*f) Tool-building:* Over the past 15 years research groups studying automated tracing have engaged individually and, eventually, jointly [17] in tool-building. Interestingly enough, for the vast majority of the tools, until very recently (when TraceLab [17] allowed for direct incorporating of existing Machine Learning libraries), the tool-building efforts proceeded in isolation from the tool- and library-building efforts in the KDD community. This, in part, can be explained by both (a) lack of reliable KDD libraries available in mid-2000s when automated tracing tool-building commenced, and (b) the need for UI/UX that is custom-tailored to the tracing problem.

*g) Automation alone is not enough:* There are two ways to produce trace relationship. First, it can be done as a byproduct of the software lifecycle, where software engineers create and maintain up-to-date traces between a variety of artifacts they build. Alternatively, and often observed in practice, tracing is done *aposteriori*, as the means of recovering a relationship that was not created or maintained during the software lifecycle. Additionally, tracing is done as part of the Verification & Validation process. In this latter scenario the presence of trace relations is often ignored as they are recovered from scratch and then compared to the existing ones. The V&V process is often used for mission- and safety-critical systems, where the cost of human error is high.

When validating the trace relationships as part of the V&V process, the candidate trace relationships produced by the automated methods must be further validated by human analysts. This turns out to be a difficult problem by itself. Our research shows that humans do not necessarily improve high accuracy candidate traces provided to them by automated methods, but may drastically improve some low accuracy ones [8, 11, 9].

*h) What if no one comes?:* KDD techniques are routinely productized in a wide range of commercial applications. However, as widely as these techniques are applied to tracing, the successful use of the KDD technologies for tracing in actual industry projects is rare.

---

[2]Given an ordered list of *n* results, the average precision of the list is the average of the precisions of lists of 1, 2,..., n results. Mean Average Precision is the average of the Average Precisions over multiple queries.

[3]The Receiver Operating Curve (ROC curve) is a mapping of a method's true positive rate vs. false positive rate plotted as a 2 Ddiagram. The area under the ROC curve is a good measure of the accuracy of a KDD technique.

[4]Which is often not known because many IR problems do not have exhaustive ground truth, making it impossible to compute recall.

[5]https://archive.ics.uci.edu/ml/index.php

[6]https://trec.nist.gov/data.html

[7]http://www.kaggle.com

## III. MOVING FORWARD

Based on our observations and lessons learned, we posit:

*a) Requirements tracing is essentially applied KDD:* The tracing community has adopted and internalized the KDD view of tracing, evaluating the artifacts, and analyzing the quality of trace relations.

*b) We need standards for ground truth:* The RE community should come to an acceptable standard of what constitutes sufficient evidence of validity of ground truth developed for tracing and other RE datasets. This will streamline both reporting and reviewing of a large number of papers, and will encourage more research groups to develop their own datasets complete with newly obtained ground truth.

This problem is well-recognized in the community, and efforts to address are underway. One such effort, the MIDAS tool, concentrates on the production of ground truth for trace relations via crowdsourcing [16].

*c) We need more datasets:* Related to the observation above, in order to progress and improve the accuracy of automated trace recovery techniques, the requirements tracing community needs more data. Development of new tracing datasets – both small and large, is an important challenge for all research groups working in this area.

This too, is a well-understood problem that exists in the larger context of the dearth of good datasets for empirical SE research. One of the ways to address this is creation of venues that welcome dataset submission – such as the PROMISE conference, and the RE Data Track. These efforts are in line with the ways by which the KDD community builds its repository of datasets.

*d) Requirements Engineering (and Software Engineering) researchers should get Data Science training:* As evidenced by the proliferation of conferences such as ASE, MSR, and PROMISE, the use of KDD methods in Software Engineering is not limited to tracing, nor is tracing the most prominent application of such methods. In the early 2000s, research in the area of applications of KDD techniques to Software Engineering almost invariably involved researchers with primary expertise in KDD collaborating with empirical Software Engineering researchers. Today, Software Engineering researchers should learn, as part of their base education, the KDD techniques and their use for data analysis. Knowing how to apply these techniques to analysis of software artifacts is no longer optional for empirical Software Engineering researchers.

## ACKNOWLEDGEMENTS

The work of the second author has been supported in part by NSF grants CCF-1511117 and CICI 1642134.

## REFERENCES

[1] G. Antoniol, G. Canfora, G. Casazza, A. De Lucia, and E. Merlo. Recovering code to documentation links in oo systems. In *Proceedings, Sixth Working Conference on Reverse Engineering*, pages 136–144, October 1999.

[2] D.M. Berry. Evaluation of tools for hairy requirements and software engineering tasks. In *Proceedings International Requirements Engineering Conference Workshops (RE'2017 Workshops)*, pages 284–291, September 2017.

[3] David M. Blei, Andrew Y. Ng, and Michael I. Jordan. Latent dirichlet allocation. *Journal of Machine Learning Research*, 3:993–1022, 2003.

[4] Markus Borg, Per Runeson, and Anders Ardö. Recovering from a decade: a systematic mapping of information retrieval approaches to software traceability. *Empirical Software Engineering*, 19(6):1565–1616, 2014.

[5] J. Cleland-Huang, O. Gotel, J.H. Hayes, P. Mäder, and A. Zisman. Software traceability: Trends and future directions. In *Proceedings, International Conference of Software Engineering (ICSE'2014)*, 2014.

[6] Jane Cleland-Huang, Brian Berenbach, Stephen Clark, Raffaella Settimi, and Eli Romanova. Best practices for automated traceability. *IEEE Computer*, 40(6):27–35, 2007.

[7] Jane Cleland-Huang, Olly Gotel, and Andrea Zisman, editors. *Software and Systems Traceability*. Springer, 2012.

[8] David Cuddeback, Alex Dekhtyar, and Jane Huffman Hayes. Automated requirements traceability: The study of human analysts. In *RE 2010, 18th IEEE International Requirements Engineering Conference, Sydney, New South Wales, Australia, September 27 - October 1, 2010*, pages 231–240, 2010.

[9] David Cuddeback, Alex Dekhtyar, Jane Huffman Hayes, Jeff Holden, and Wei-Keat Kong. Towards overcoming human analyst fallibility in the requirements tracing process. In *Proceedings of the 33rd International Conference on Software Engineering, ICSE 2011, Waikiki, Honolulu , HI, USA, May 21-28, 2011*, pages 860–863, 2011.

[10] Scott C. Deerwester, Susan T. Dumais, Thomas K. Landauer, George W. Furnas, and Richard A. Harshman. Indexing by latent semantic analysis. *JASIS*, 41(6):391–407, 1990.

[11] Alex Dekhtyar, Olga Dekhtyar, Jeff Holden, Jane Huffman Hayes, David Cuddeback, and Wei-Keat Kong. On human analyst performance in assisted requirements tracing: Statistical analysis. In *RE 2011, 19th IEEE International Requirements Engineering Conference, Trento, Italy, August 29 2011 - September 2, 2011*, pages 111–120, 2011.

[12] Tom Fawcett. An introduction to ROC analysis. *Pattern Recognition Letters*, 27(8):861–874, 2006.

[13] O. Gotel and A. Finkelstein. An analysis of the requirements traceability problem. In *Proceedings of the 1st IEEE International Conference on Requirements Engineering*, pages 94 —-101, April 1994.

[14] Jane Huffman Hayes, Alex Dekhtyar, Senthil Karthikeyan Sundaram, and Sarah Howard. Helping analysts trace requirements: An objective look. In *12th IEEE International Conference on Requirements Engineering (RE 2004), 6-10 September 2004, Kyoto, Japan*, pages 249–259, 2004.

[15] J.H. Hayes, A. Dekhtyar, and S.K.Sundaram. Advancing candidate link generation for requirements tracing: The study of methods. *IEEE Transactions of Software Engineering*, 32(1):4 – 19, 2006.

[16] Albert Kalim, Satrio Husodo, Jane Huffman Hayes, and Erin Combs. Multi-user input in determining answer sets (MIDAS). In *Proceedings of the International Conference on Requirements Engineering (RE'2018)*, August 2018.

[17] E. Keenan, A. Czauderna, G. Leach, J. Cleland-Huang, Y. Shin, E. Moritz, M. Gethers, D. Poshyvanyk, J. Maletic, J. H. Hayes, A. Dekhtyar, D. Manukian, S. Hossein, and D.Hearn. Tracelab: An experimental workbench for equipping researchers to innovate, synthesize, and comparatively evaluate traceability solutions. In *Proceedings, 34th International Conference on Software Engineering (ICSE'2012*, pages 1375–1378, June 2012.

[18] A. Marcus and J. I. Maletic. Recovering documentation-to-source-code traceability links using latent semantic indexing. In *25th International Conference on Software Engineering, 2003. Proceedings.*, pages 125–135, May 2003.
[19] Rocco Oliveto, Malcom Gethers, Denys Poshyvanyk, and Andrea De Lucia. On the equivalence of information retrieval methods for automated traceability link recovery. In *The 18th IEEE International Conference on Program Comprehension, ICPC 2010, Braga, Minho, Portugal, June 30-July 2, 2010*, pages 68–71, 2010.
[20] Stephen E. Robertson and Karen Spärck Jones. Relevance weighting of search terms. *JASIS*, 27(3):129–146, 1976.
[21] Stephen E. Robertson, C. J. van Rijsbergen, and Martin F. Porter. Probabilistic models of indexing and searching. In *Proc. Joint ACM/BCS Symposium in Information Storage and Retrieval*, pages 35–56, 1980.
[22] J.J. Rocchio. Relevance feedback in information retrieval. In *The SMART Retrieval System: Experiments in Automatic Document Processing*, pages 313–323. Prentice Hall, 1971.
[23] Stefan Winkler and Jens von Pilgrim. A survey of traceability in requirements engineering and model-driven development. *Software and System Modeling*, 9(4):529–565, 2010.